\documentclass[twocolumn,showpacs,preprintnumbers,amsmath,amssymb,prl]{revtex4-1}
\usepackage{color}

\usepackage{graphicx}
\usepackage{dcolumn}
\usepackage{bm}
\usepackage{subfigure}
\usepackage{amssymb}
\usepackage{tabularx}
\usepackage{multirow}

\begin{document}

\title{Oscillation Spectrum of a Magnetized Strongly Coupled One-Component-Plasma}

\author{T.~Ott$^1$}
\author{H.~K\"ahlert$^1$}
\author{A.~Reynolds$^{1,2}$}
\author{M.~Bonitz$^1$}%
\affiliation{$^1$Christian-Albrechts-Universit\"at zu Kiel, Institut f\"ur Theoretische Physik und Astrophysik, Leibnizstra\ss{}e 15, 24098 Kiel, Germany \\
$^2$School of Physics and Astronomy, University of Birmingham, Edgbaston, Birmingham B15 2TT, UK}

\date{\today}

\begin{abstract}
A first-principle study of the collective oscillation spectrum of a strongly correlated one-component plasma in a strong magnetic field is presented. The spectrum consists of six fundamental modes 
which are found to be in good agreement with results from the Quasi-Localized Charge approximation. At high frequencies additional modes are observed that include Bernstein-type oscillations and their transverse counterparts which are of importance for the high-frequency optical and transport properties of these plasmas.
\end{abstract}

\maketitle


Strongly coupled charged particle systems in which the interaction energy exceeds the kinetic energy are of rapidly growing 
interest in many fields reaching from condensed matter to ultracold gases, trapped ions, nonideal plasmas up to the quark-gluon plasma, e.g. \cite{Bonitz2010a}. Quite often these systems are subject to a strong magnetic field -- examples are white dwarf stars, neutron star crusts~\cite{Peng2007,baiko09}, magnetized target fusion scenarios~\cite{Basko2000,Deutsch2009} or quantum Hall systems \cite{klitzing_2002}. Magnetic field effects are also coming into the focus of laboratory experiments with dusty plasmas~\cite{Bonitz2010a}, ultracold neutral plasmas~\cite{Killian2007} and trapped ions \cite{dantan10}. Apart from basic transport properties  such as diffusion~\cite{Ott2011c}, the collective excitation spectra are of fundamental relevance e.g. for the response to optical excitation.
In two-dimensional (2D) plasmas, the properties of these wave spectra are firmly established analytically via the Quasi-Localized 
Charge Approximation (QLCA) of Kalman and Golden~\cite{Golden1993,kalman05,Jiang2007} and computer simulations~\cite{Hou2009c}. Strong correlations lead to the existence of a shear-type mode~\cite{Donko2008,Upadhyaya2011} alongside the familiar compressional modes (plasmons) known from weakly coupled situations. These shear modes could be verified in 2D dusty plasma experiments \cite{Pieper1996,Piel2006}.
In the presence of a strong magnetic field, these transform into two modes commonly known as 
magnetoplasmon and -shear or upper and lower hybrid mode, e.g. \cite{uchida2004}. 

Surprisingly, the corresponding behavior of the much more common strongly coupled {\em three-dimensional plasmas} has remained unexplored, except  for a preliminary study in Ref.~\cite{Kalman1994}. The present work is, therefore, devoted to filling this gap by a theoretical approach based on the QLCA and complementary first-principle molecular dynamics (MD) simulations. We obtain the complete set of fundamental linear oscillations and explore their wave-vector dispersion and polarization. Furthermore we present evidence for the existence of nonlinear modes that are generalizations of Bernstein modes of high-temperature plasmas \cite{bernstein58,Bellan2006}.

\begin{figure}
\centering
\includegraphics{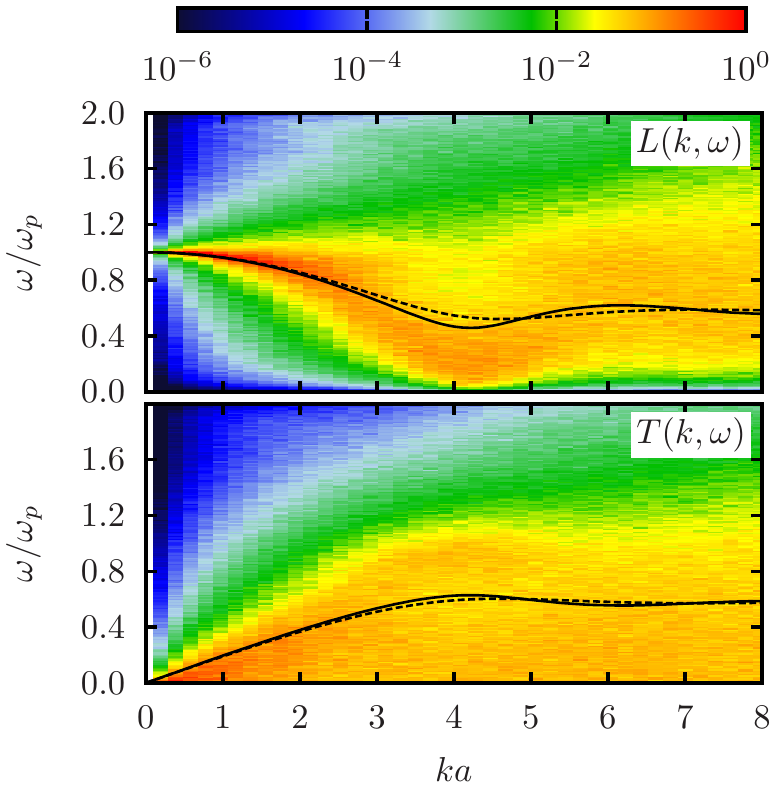}
\caption{\label{fig:3d_spec_k0g100b0} (color) Longitudinal spectrum $L(k,\omega)$ and transverse spectrum $T(k,\omega)$ 
of an unmagnetized {OCP} at $\Gamma=100$. The black lines are the {QLCA} dispersion 
relations (Plasmon and ordinary shear mode) for $\Gamma=100$ (solid line) and $\Gamma=20$ (dashed line). 
}
\end{figure}

{\bf Model and simulation method.} We consider a classical one-component plasma (with a neutralizing homogeneous background, OCP) subject to an external magnetic field ${\bf B} = B\hat{\bf e}_z$. The exact equations of motion for $N$ particles are
\begin{equation}
\ddot{\bf r}_i = {\bf F}_i/m + \omega_c\,{\bf v}_i \times \hat{\bf e}_z, \quad i=1\dots N,
\label{eq:eom}
\end{equation}
where ${\bf v}_i=\dot{\bf r}_i$ and $F_i$ is the Coulomb force due to all particles $j\neq i$. 
The thermodynamic equilibrium state of the system is characterized by the coupling parameter~$\Gamma=q^2\times (4\pi\varepsilon_0 ak_BT)^{-1}$, 
where $a=\left[3/\left (4 n \pi \right)  \right]^{1/3}$ is the Wigner-Seitz-radius with the number density $n$, and $q$ denotes the particle charge. The strength of the magnetic field $B$ is given by $\beta=\omega_c/\omega_p \propto B$, i.e., the ratio of the cyclotron frequency, $\omega_c = qB/m$, and the plasma frequency, $\omega_p^2= n q^2 /\left( \varepsilon_0 m\right)$.
The spectra of the longitudinal and transverse fluctuations, $L(k,\omega)$ and $T(k,\omega)$, of the OCP follow from the Fourier component of the general current density operator in standard manner~\cite{Hansen1975,Boon1991}
\begin{align}
 {\bf j}(k, t) = \sum_{j=1}^N {\bf v}_j(t) \exp[{i {\bf k} \cdot {\bf r}_j(t)}],
 \label{eq:j_b=0}
\end{align}
 via a temporal Fourier transform $\mathcal F_t$ 
\begin{align}
& \frac{1}{2\pi N} \lim_{t\rightarrow\infty} \frac{1}{t} \left \vert \mathcal F_t\{{\bf j}(k, t)\}\right \vert^2.
\label{eq:ft}
\end{align}
The current can be decomposed into a part parallel and a part perpendicular to ${\bf k}$, ${\bf j}={\bf j}_{\parallel} + {\bf j}_{\perp}$, so application of (\ref{eq:ft}) to ${\bf j}_{\parallel}$ yields the longitudinal spectrum $L(k,\omega)$ and ${\bf j}_{\perp}$ the transverse spectrum 
$T(k,\omega)$. Collective oscillations appear as peaks in these spectra, cf. Fig.~\ref{fig:3d_spec_k0g100b0}.

{\bf Zero magnetic field}. To set the stage, consider first the unmagnetized system. In Fig.~\ref{fig:3d_spec_k0g100b0} we plot the longitudinal and transverse and longitudinal fluctuation spectra $L(k,\omega)$ and $T(k,\omega)$ together with the well-known QLCA results for the collective modes (solid black lines~\cite{Kalman1990,Donko2008}). There are two remarkable deviations from the familiar spectrum of a weakly coupled plasma, e.g. \cite{alexandrov88,Bellan2006}: first, the plasmon dispersion does not show a monotonic increase as $\omega^2(k)=\omega_p^2(1+3k^2 r_D^2)$ [$r_D$ is the Debye radius] but decays and exhibits oscillations. Second, there exists an additional shear mode in the transverse spectrum.  Fig.~\ref{fig:3d_spec_k0g100b0} also shows that the precise value of $\Gamma$ is of minor importance, increase of the coupling  tends to slightly increase the amplitude of the oscillations. 
Overall, Fig.~\ref{fig:3d_spec_k0g100b0} indicates that QLCA describes the oscillation spectrum of a strongly coupled OCP rather well \cite{note_2peaks}.

\begin{figure}
\centering
\includegraphics{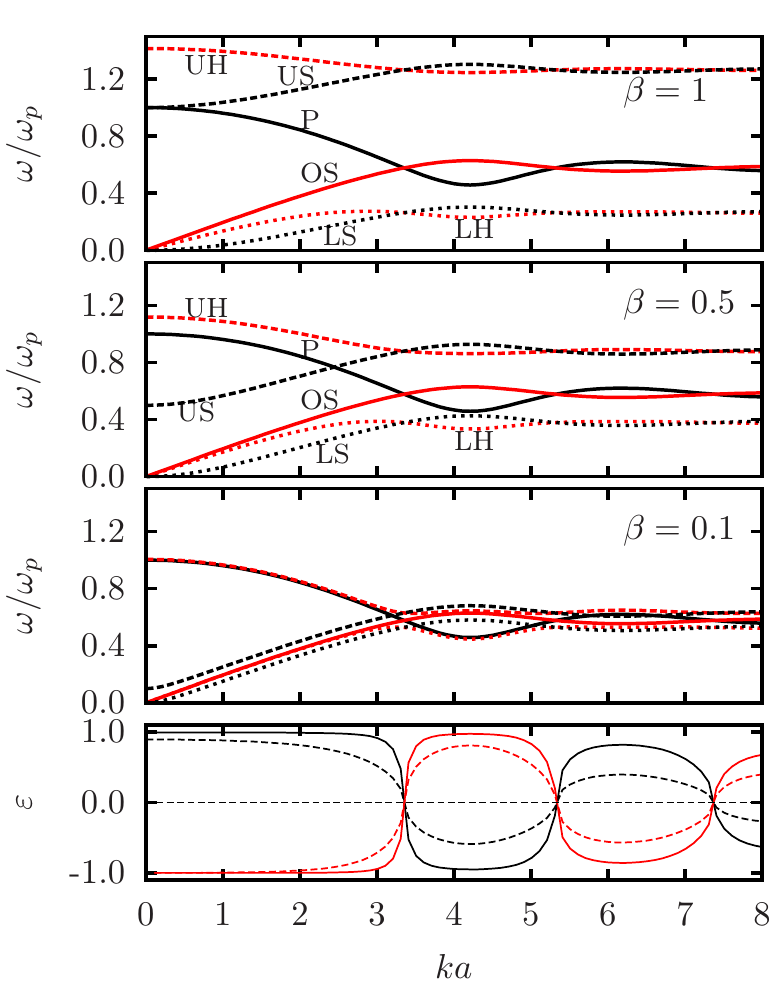}
\caption{\label{fig:qlca_3d_b_k0} (color online) The modes of a 3D magnetized OCP ($\kappa=0$, $\Gamma=100$) in QLCA, Plasmon (P), Ordinary shear (OS), upper hybrid (UH), lower hybrid (LH), lower shear (LS) and upper shear (US), for three magnetic field strengths. Modes with 
${\bf k} \parallel {\bf B}$ are shown in black (dark), and those with ${\bf k} \perp {\bf B}$ in red (light). Bottom: $k$-dependence of the eccentricity $\varepsilon$ of the UH (black) and LH (red) for $\beta=0.1$ (solid) and $\beta=0.5$ (dashed).
}
\end{figure}

{\bf Theory for a magnetized OCP}. We now turn to a theoretical description of the wave spectrum in a magnetized strongly correlated OCP using the QLCA approach, e.g.~\cite{Kalman1990,Golden1993,Jiang2007}. The equation of motion for the collective coordinates $\eta_{{\bf k}\alpha}(\omega)$ in the presence of a magnetic field read 
\begin{eqnarray}
 \Big[ \omega^2\delta_{\alpha\beta} -D_{\alpha\beta}({\bf k}) - \frac{k_\alpha k_\beta}{k^2}\omega_p^2 +i\omega\omega_c\sigma_{\alpha\beta}\Big]\eta_{{\bf k}\alpha}(\omega) = 0, \quad
\label{eq:qlca_eom}
\\
  \mbox{where}\qquad\sigma = \left( {\begin{array}{ccc}
 0&-1&0  \\
 1&0&0  \\
 0&0&0\\
 \end{array} } \right) ,\nonumber
\end{eqnarray}
and $D_{\alpha\beta}$ is the QLCA dynamical matrix~\cite{Golden2000} which is a functional of the pair distribution function $g(r)$.  
The non-trivial solutions to Eq.~\eqref{eq:qlca_eom} represent the modes of the system which turn out be rather complicated. Therefore, here 
 we concentrate on modes with ${\bf k} \parallel {\bf B}$ or ${\bf k} \perp {\bf B}$. Their main properties are summarized in Tab.~\ref{tab:qlca_modes} and the dispersions are depicted for three different magnetic field strengths at strong coupling ($\Gamma=100$) in Fig.~\ref{fig:qlca_3d_b_k0}. 
Obviously, there exist five possible (pairwise parallel or perpendicular) orientations of the vectors ${\bf B}, {\bf k}$ and ${\bf r}$ -- two (three) of them corresponding to ${\bf k} \parallel {\bf B}$ (${\bf k} \perp {\bf B}$), cf. Tab.~\ref{tab:qlca_modes}. At the same time, the QLCA equation (\ref{eq:qlca_eom}) yields six fundamental solutions -- three for each orientation of the wave vector -- which we discuss in the following.
\begin{table*}[ht]
\centering
 \renewcommand{\arraystretch}{1.5}
  \begin{tabularx}{\textwidth}{l|c|c|c|c|c|c|c|c} 
    {Name}    & $\vec k\angle \vec B$& $\vec r\angle \vec B$& $\vec k\angle \vec r$&{Type} &{Polarization} & {Dispersion $\omega_i(k)$ } & {Asymptotic} $\omega_i^{\infty}$ & Spectrum\\   
\hline\hline
Plasmon (P)&  $\parallel $ &$\parallel $ &$\parallel $ & longit. & --- & $\sqrt{\omega^2_p + D_L(k)}$ & $\omega_E$ &$L^{\parallel}$\\[1ex]
Upper Shear (US)&$\parallel $&$\perp $&$\perp $& transv. & circular &  $\frac{1}{2}\Big[ \sqrt{\omega_c^2 + 4 D_T(k)} + \omega_c\Big]$  & $\frac{1}{2}\Big[\sqrt{\omega_c^2 + 4 \omega_E^2} + \omega_c\Big]$ & $T^{\perp}_{\parallel}$\\[1ex]
Lower Shear  (LS)&$\parallel $&$\perp $&$\perp $& transv.  & circular &  $\frac{1}{2}\Big[ \sqrt{\omega_c^2 + 4 D_T(k)} - \omega_c\Big]$ & $\frac{1}{2}\Big[\sqrt{\omega_c^2 + 4 \omega_E^2} - \omega_c\Big] $ & $T^{\perp}_{\parallel}$\\[1ex]\hline
Ordinary Shear (OS)&  $\perp $&$\parallel $&$\perp $  & transv. & linear &$\sqrt{D_T(k)}$ & $\omega_E$ & $T^{\parallel}$\\[1ex]
Upper Hybrid (UH)&  $\perp $&$\perp $&t-dep.  & hybrid & in-plane elliptical &$\frac{1}{\sqrt 2} \Big[ \Omega_L^2 + D_T(k) + \Omega_T^2\Big]^\frac{1}{2}$& $\frac{1}{2}\Big[\sqrt{\omega_c^2 + 4 \omega_E^2} + \omega_c\Big]$ & $L^{\perp}+T^{\perp}_{\perp}$\\[1ex]
Lower Hybrid (LH)&  $\perp $&$\perp $&t-dep.  & hybrid & in-plane elliptical &$\frac{1}{\sqrt 2} \Big[ \Omega_L^2 + D_T(k) - \Omega_T^2\Big]^\frac{1}{2}$& $\frac{1}{2}\Big[\sqrt{\omega_c^2 + 4 \omega_E^2} - \omega_c\Big]$ & $L^{\perp}+T^{\perp}_{\perp}$\\
  \end{tabularx}
  \caption{The six principal collective modes in a magnetized three-dimensional OCP in QLCA. Shown are the relative 
orientations of the wave vector $\vec k$, magnetic field $\vec B$, and particle displacement $\vec r$, the wave type, the polarization, 
the dispersion relation $\omega_i(k)$, its asymptotic, $\omega_i^{\infty}$, for large $k$ and the current spectrum in which they appear. $\omega_E=\omega_p/\sqrt{3}$ is the Einstein frequency, $D_{L/T}$ denote the longitudinal and transverse parts of $D_{\alpha\beta}$,  $\Omega_T^4 = \Big[\Omega_L^2 - D_T(k)\Big]^2 + 4\omega_c^2 D_T(k)$ and $\Omega_L^2= \omega_p^2 + \omega_c^2 + D_L(k)$.}
  \label{tab:qlca_modes}
\end{table*}

{\bf Wave vector parallel to B:} The three solutions of this type (black lines in Fig.~\ref{fig:qlca_3d_b_k0}) are the longitudinal plasmon (P) which is unaffected by the magnetic field, since the particles oscillate parallel to ${\bf B}$,
and two shear-mode solutions (upper and lower shear, US, LS). The latter arise from the shear mode of an unmagnetized plasma, cf. Fig.~\ref{fig:3d_spec_k0g100b0}, whose degeneracy is lifted by the magnetic field.  Their frequency difference equals $\omega_c$ (cf. Tab.~\ref{tab:qlca_modes}) and thus grows linearly with $B$. 

{\bf Wave vector perpendicular to B:} The three solutions (red lines in Fig.~\ref{fig:qlca_3d_b_k0}) are 
the ordinary shear mode (OS) and the upper and lower hybrid modes (UH and LH). 
The ordinary shear mode is $B$-independent since the particle oscillation occurs along ${\bf B}$. The long-wavelength limit of the UH (LH) mode  is $\omega=(\omega_p^2+\omega_c^2)^{1/2}$ ($\omega=0$). These two modes have peculiar polarization properties. While all four other modes are characterized by a fixed orientation of the displacement vector relative to ${\bf k}$ -- there are one longitudinal (P) and three transverse oscillations (US, LS, OS), the UH and LH modes exhibit particle oscillations that rotate with time
elliptically in the plane perpendicular to ${\bf B}$. To clarify these properties, the bottom part of Fig.~\ref{fig:qlca_3d_b_k0} shows the modified eccentricity $\varepsilon=\gamma\sqrt{1-b^2/a^2}$, where the semi-major axis $a=\max\big\{\vert\eta_x\vert,\vert\eta_y\vert\big\}$ 
and the semi-minor axis $b=\min\big\{\vert\eta_x\vert,\vert\eta_y\vert\big\}$ and $\gamma$ equals $1$ if $\eta_x>\eta_y$ and $-1$ otherwise. 
Here, $\vec \eta$ is the eigenvector of Eq.~\eqref{eq:qlca_eom} for the solutions UH and LH, respectively. $\varepsilon=+1$ ($-1$) for longitudinal (transverse) waves. 
Moreover, the shape of this ellipse is $k$-dependent, cf. Fig.~\ref{fig:qlca_3d_b_k0} \cite{note_uchida}. The UH mode starts longitudinal, at long wavelengths, and transforms into a predominantly transverse wave, at the first crossing of the P and OS modes. At each subsequent crossing point, the dominant character of the wave changes again. The LH mode exactly mirrors this behaviour. For $k\rightarrow\infty$, both modes are in-plane circular. 
The limit $B\rightarrow 0$ of these modes is particularly interesting: while all 6 modes converge to the two modes shown in Fig.~\ref{fig:3d_spec_k0g100b0}, the LH and UH modes only do so in a pointwise manner. The LH, at all $k$, remains below the UH, i.e. we observe an avoided crossing of the two at the intersection points of the plasmon and shear mode (see the case $\beta=0.1$ in Fig.~\ref{fig:qlca_3d_b_k0}).

\begin{figure}
\centering
\includegraphics{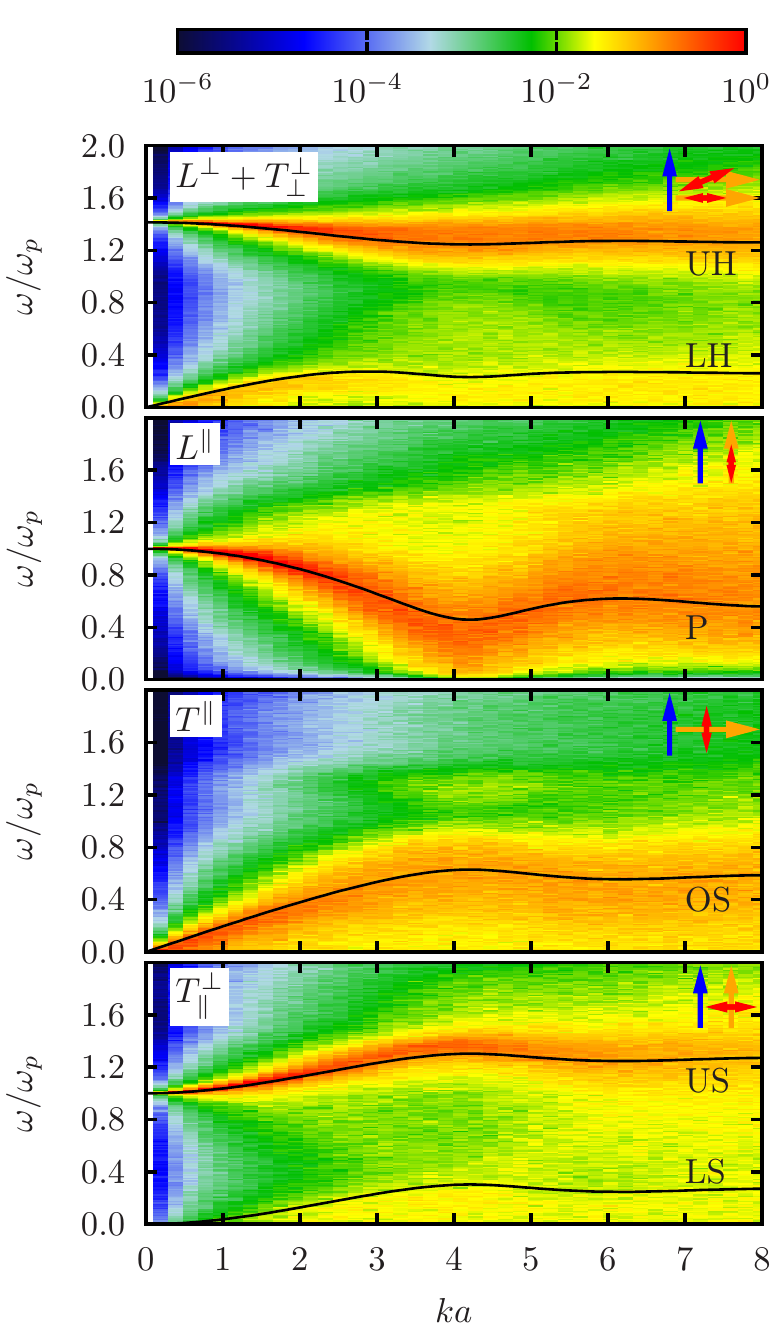}
\caption{\label{fig:3d_spec_k0g100b10_4} (color) The complete set of oscillation spectra of a 
magnetized OCP at $\Gamma=100$ and $\beta=1.0$ for frequencies $\omega \le 2\omega_p$. The solid lines 
correspond to the six {QLCA} modes that are labeled in the figure, cf. Tab.~\ref{tab:qlca_modes}.
The orientation of the magnetic field (blue), wave vector (orange) and the particle oscillations (red) are indicated by the arrows.  
}
\end{figure}

\begin{figure*}
\centering
\includegraphics{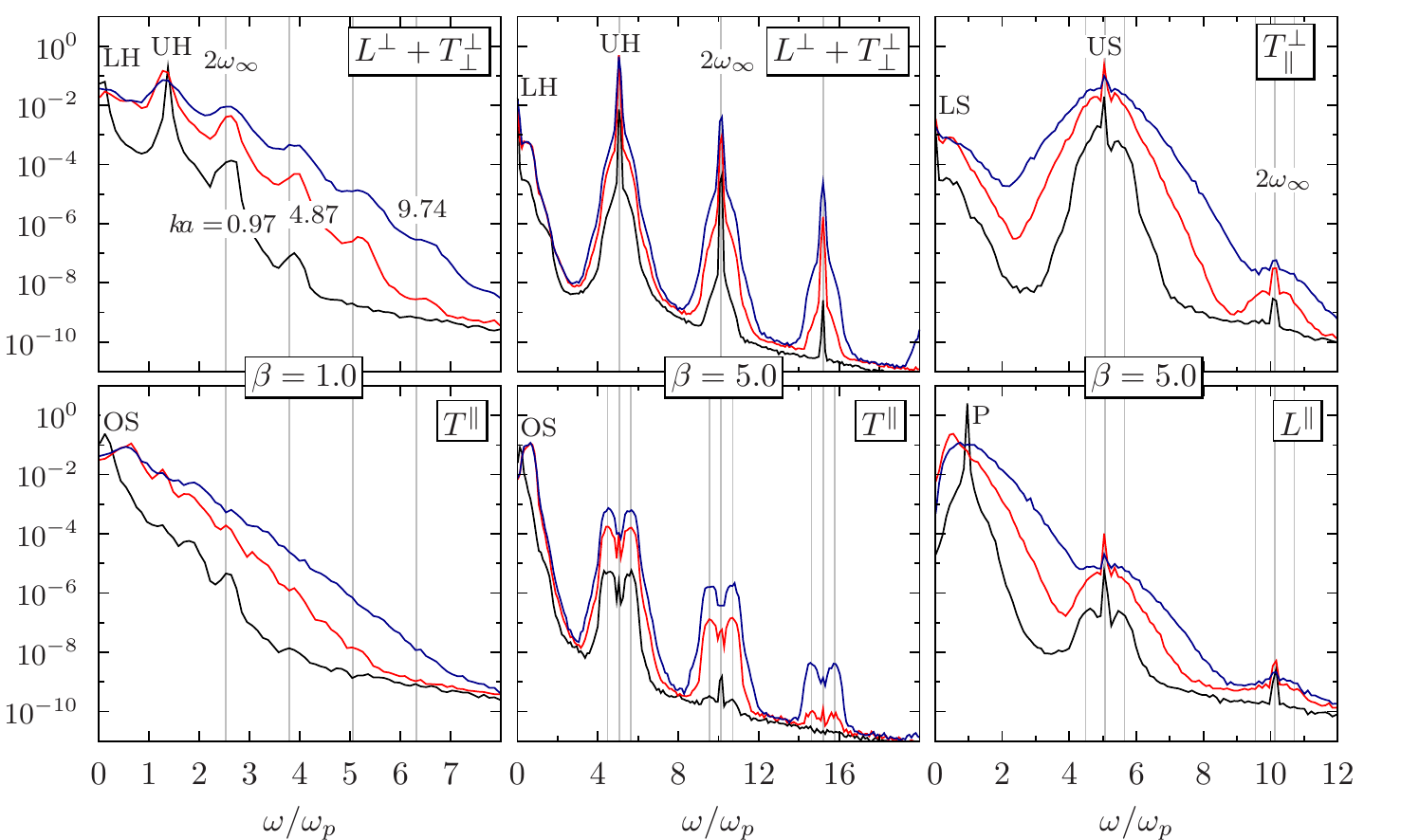}
\caption{\label{fig:3d_hhg} (color online) Fluctuation spectra of an OCP extended to high frequencies, at two values of the magnetic field and three wave numbers, as indicated in the figure. Vertical lines indicate the harmonics $n\omega_{\infty}$ and the frequencies  
$n \omega_{\infty}\pm{\omega_E}$.}
\end{figure*}

{\bf Simulation results}. Let us now critically test the theoretical predictions against the computer experiment. 
To this end, Eq.~\eqref{eq:eom} is solved by a microcanonical MD method for $N=8000$ particles, incorporating an arbitrarily strong static homogeneous magnetic field into the second-order Velocity Verlet algorithm \cite{Spreiter1999}. Prior to data acquisition, the system is equilibrated by an 
isokinetic thermostat. The above mentioned five possible orientations of ${\bf B}, {\bf k}$ and ${\bf r}$ manifest themselves in five different currents of the type (\ref{eq:j_b=0}) leading to five fluctuation spectra: there are two longitudinal ones, $L^{\parallel}$ and $L^{\perp}$ [the superscript denotes the mutual orientation of ${\bf B}$ and ${\bf r}$] and three transverse ones, $T^{\parallel}$, $T^{\parallel}_{\parallel}$ and $T^{\parallel}_{\perp}$ [the subscript specifies the angle between ${\bf B}$ and ${\bf k}$ in case of an ambiguity]. The correspondence of the five fluctuation spectra to the six collective oscillations can be seen in Tab.~\ref{tab:qlca_modes}.

Fig.~\ref{fig:3d_spec_k0g100b10_4} shows the wave spectra of a magnetized OCP at $\Gamma=100$ and $\beta=1.0$. To capture the two hybrid modes, the two spectra $L^{\perp}(k,\omega)$ and $T^{\perp}_{\perp}(k,\omega)$ have been added together. Six frequency peaks 
can be identified which overall show good agreement with the QLCA modes. This conclusion is representative and holds also for other values of $\Gamma$ and $\beta$.  At the same time, there exist three noticeable deviations of the QLCA from the simulations. The first is observed at low frequencies (corresponding to long times) due to the breakdown of the QLCA assumption of a frozen potential landscape \cite{kalman05}. Second, the low-frequency shear modes vanish at small $k$, which is not reproduced in QLCA \cite{ohta00,murillo00}. Finally, it has been observed that QLCA misses the high-frequency part of the spectrum in a 2D magnetized OCP where Bernstein-type modes were found in the simulations \cite{Bonitz2010,Ott2011}. It is, therefore, tempting to analyze more in detail the frequency region that is not captured by Fig.~\ref{fig:3d_spec_k0g100b10_4}. 

In the left column of Fig.~\ref{fig:3d_hhg} we present the high-frequency part of the spectra corresponding to ${\bf B}\perp {\bf k}$ at three values of $k$. In addition to the two known modes (UH, LH), the spectrum $L^{\perp}+T_{\perp}^{\perp}$ exhibits a number of further peaks that are equally spaced. While these peaks are strongly damped for $\beta=1$, they become very sharp at a five-fold stronger magnetic field, cf. center column of Fig.~\ref{fig:3d_hhg}. These modes in the spectrum of waves propagating perpendicular to the magnetic field clearly resemble Bernstein modes of high-temperature plasmas, e.g. \cite{bernstein58,Bellan2006}. In a 2D OCP it was found recently that ``dressed'' (modified by correlations) Bernstein modes reappear at strong coupling \cite{Bonitz2010,Ott2011}.
The analysis of the present 3D spectra shows that these modes appear at multiples of the high-$k$ limit of the upper shear and upper hybrid frequency (see the asymptotics in Tab.~\ref{tab:qlca_modes}), 
$\omega^\infty_{UH} = \omega^\infty_{US} \equiv \omega_\infty = \frac{1}{2}[(\omega_c^2 + 4\omega_E^2)^{1/2}+\omega_c]$, with the Einstein frequency $\omega_E=\omega_p/\sqrt{3}$. 

Apart from this similarity to a 2D OCP, the high-frequency spectrum in 3D is much richer, and the harmonics $n\omega_\infty$ show up also in other fluctuations. Indeed, as can be seen in the lower left part of Fig.~\ref{fig:3d_hhg} the harmonics emerge also in the 
spectrum $T^{\parallel}$ around the same frequencies (solid grey lines) and more clearly at stronger fields, cf. bottom center figure. A closer inspection shows that there these peaks are visible only at small $ka$ whereas, at higher $ka$, instead, a double peak structure emerges at $n \omega^\infty\pm \omega_E$. 
This is accompanied by a broadening of the higher harmonics peaks in $L^{\perp}(\omega)+T_{\perp}^{\perp}(\omega)$ that
develop ``shoulders'' around $n \omega_\infty\pm \omega_E$. A similar trend is observed
in the two ${\bf B} \parallel {\bf k}$ excitation spectra, cf. rightmost column of Fig~\ref{fig:3d_hhg}.
Here, the US mode is clearly seen in $T^{\perp}_{\parallel}(\omega)$, as a peak around $\omega/\omega_p \approx 5$. 
Additionally, the second harmonic of the US is visible around $\omega/\omega_p \approx 10$. 
These two peaks also appear in $L^{\parallel}(\omega)$, alongside the low-frequency plasmon peak at 
$\omega/\omega_p\approx 1/\sqrt{3}$, cf. Tab.~\ref{tab:qlca_modes}. 
The peaks at $\omega/\omega_p \approx 5,10$ are accompanied by the same double peak structure seen in the cross-field wave spectra. 

All these high-frequency features can be understood within a common simple physical picture: {\em scattering of multiple 
short wavelength oscillations} where $\omega_1 + \omega_2 = \omega_{12}$ and ${\bf k}_1 + {\bf k}_2 = {\bf k}_{12}$, a mechanism well-known in high-temperature plasmas \cite{alexandrov88,Bellan2006} and nonlinear optics. In fact, the $n$-th harmonic of the US (UH) is explained by a cascade of nonlinear inelastic interactions $\omega_\infty \rightarrow 2\omega_\infty \rightarrow \dots n\omega_\infty$. On the other hand, the pair of side peaks at 
$\omega_\infty\pm \omega_E$ in the middle column of Fig.~\ref{fig:3d_hhg} arise from the scattering of an UH (frequency $\omega_\infty$) with an OS wave (frequency $\omega_E$), and the analogous peaks in the right column of Fig.~\ref{fig:3d_hhg} are due to a scattering of an US mode and a plasmon (frequencies $\omega_\infty$ and $\omega_E$). Finally, the appearance of the peaks at $\omega/\omega_p \approx 5,10,15$ in $T^\parallel$ [at $\omega/\omega_p \approx 5, 10$ in $L^\parallel$] are due to {\em elastic scattering} of an UH [US] mode (or a corresponding higher harmonic) with that frequency on the particles during which the particle oscillation direction is turned from in-plane ($\perp {\bf B}$) to out-of plane ($\parallel {\bf B}$) which requires a finite shear elasticity -- a remarkable feature of strongly correlated plasmas \cite{diff_comment}.
It is particularly striking that this mechanism also excites novel longitudinal modes in the plasmon spectrum $L^\parallel$ which have a $B$-dependent frequency, in contrast to the fundamental plasmon.

In summary, we have presented the complete collective oscillation spectrum of a strongly correlated one-component plasma in a magnetic field. The OCP supports six base modes -- three for each ${\bf k}$ -- which are well described by the QLCA, and their characteristics have been given in Tab.~\ref{tab:qlca_modes}.
In addition to these modes there exist additional oscillations arising from elastic and inelastic wave scattering. They are fuelled by the 
familiar Bernstein scenario of harmonics generation in a magnetic field. However, in contrast to high-temperature plasmas the base frequency differs from $\omega_c$ -- it is modified by correlations. Furthermore, correlations give rise to a novel class of modes: {\em transverse Bernstein waves} which are harmonics of the upper shear oscillation. The present results are representative for a broad range of coupling parameters and for Coulomb and Yukawa OCP as well. They are, therefore, expected to be of direct relevance for the high-frequency transport and optical response of a broad spectrum of strongly coupled magnetized plasmas -- from white dwarf and neutron stars to trapped ions and dusty plasmas.

This work is supported by the DFG via SFB-TR 24 via projects A5 and A7, the DAAD via the RISE program, and a grant 
for CPU time at the North-German Supercomputing Alliance HLRN. 


%

\end{document}